\documentclass[18pt,a4paper,final,twoside]{article}
\usepackage[utf8x]{inputenc}
\usepackage{ucs}
\usepackage[english]{babel}
\usepackage{amsmath}
\usepackage{amsfonts}
\usepackage{amssymb}
\usepackage{makeidx}
\usepackage{graphicx}
\usepackage{lmodern}

\newcommand{\Lim}{\text{Lim }}
\newcommand{\PV}{\mathcal{P}\hspace*{-0.25ex}}

\begin{document}

\begin{titlepage}

\begin{center}
{\Large On The Pauli-Weisskopf Anti-Dirac Paper.}\\
\vspace{5em}

\large
Walter Dittrich\\
\makeatletter
Institute for Theoretical Physics\\
University of T\"{u}bingen\\
Auf der Morgenstelle 14\\
D-72076 T\"{u}bingen\\
Germany\\
qed.dittrich@uni-tuebingen.de\\
\makeatother
\end{center}
\end{titlepage}

\begin{abstract}
We review in this article the role which the work of Pauli and Weisskopf played in formulating a quantum field theory of spinless particles.
To make our computations as transparent as possible, we offer a physicist's derivation of the Klein-Gordon-Fock equation. Since invariant functions play a
significant part in our paper, we will discuss them in great detail. We emphasize Pauli's and Weisskopf's view that Dirac's hole
theory is totally obsolete in formulating a consistent quantum field theory, be it for scalar or spinor particles.
As an important example we present the calculation for producing charged scalar particles in an external electric field.
\end{abstract}

\section{Introduction}
After Pauli and Weisskopf published their anti-Dirac paper\cite{Pauli1934}, the Klein-Gordon-Fock (K.G.F.) field theory became a well-respected concept
for describing the behavior of massive spinless scalar particles, like pions. 
Homogeneous solutions of the K.G.F. equation as well as Green's functions for the inhomogeneous K.G.F. equation were worked out in detail.
It might not be well known that Pauli and Majorana never thought very highly of Dirac's hole theory. 
More recently a paper was published in which Majorana, several years before Pauli and Weisskopf, studied the quantization of the relativistic K.G.F. equation\cite{Esposito2007}.

But the final blow to Dirac's hole theory came from Pauli and Weisskopf. Admittedly, it took a long time until Dirac's idea was relegated to the corner of "mere historical interest".
With the discovery of pions and other (pseudo-)scalar particles it became clear that one could do without Dirac holes for antiparticles.
Pauli and Weisskopf had always rejected the Dirac equation as an equation for a relativistic probability amplitude.
They regarded the Dirac equation as a relativistic matter field equation and not an equation of the probability amplitude in the $(x,y,z)$ space.
Although the idea of a second quantized field operator $\psi(\vec{x},t)$ for many-particle systems was clear to Jordan in the three-man paper\cite{Born1925} of 1925,
it was Pauli and Weisskopf who insisted that a concept like the probability of a particle to be found in $\vec{x}$ space does not make much sense for relativistic particles,
and this holds true for electrons, photons and K.G.F. particles alike.
Consequently, the Dirac equation as well as the K.G.F. equation should be treated as matter-field equations for many particles rather than as an equation for single-particle probability
amplitudes\cite{Tomonaga1997}.

After having constructed the relativistic scalar field theory, Pauli wondered "why nature had made no use of the possibility of the theory that there exist spin-zero bosons [...]".
Needless to say, Pauli's question was answered by nature in the affirmative. Not long after Pauli's and Weisskopf's paper of 1934, the relativistic scalar wave field theory would be finally
established in the appearance of $\pi$ mesons.

Admittedly, most of the facts in the present paper have been known for quite some time. But what is probably new is the attempt to construct the probability amplitude
$\langle\vec{x},t \vert \vec{x}_0,t_0=0\rangle$ in relativistic single-particle quantum mechanics(QM). We will discover that the probability amplitude of finding the particle
outside the light cone is not zero - it drops exponentially. So it can be found in an area which is forbidden by special relativity. To get out of this trouble we will use instead the
the language of second quantization. Here we will study the Pauli-Jordan commutator function $\left[ \psi(\vec{x},t),\psi^\dagger (\vec{x}_0)\right]$
which is related to the probability amplitude of detecting the particle at $(\vec{x},t)$ while it was created at $(\vec{x}_0,t_0=0)$.

Our first attempt ends up in a disaster. Not only have we violated causality - meaning no signal can travel faster than light. In addition, we find a breakdown of simultaneity, i.e., we cannot
give $\left[\psi,\psi^\dagger\right]$ an invariant meaning outside the light cone.

After we have remedied these failings, we begin a thorough study of the so-called invariant functions and propagation functions. Here we rely on the totally neglected but
wonderful article by J. Schwinger\cite{Schwinger1949}. Finally, we present a list of invariant functions of the K.G.F. theory in $(\vec{x},t)$ space.

\section{The Free Klein-Gordon-Fock Theory - Particle Description}
Our goal is to investigate the consequences of particles with relativistic energy spectrum ($\hbar=c=1$):
\begin{equation}
H:= \sum_{\vec{p}} a^\dagger(\vec{p})\sqrt{\vec{p}^2+m^2} a(\vec{p})\quad .\label{eq:DefHamiltonian}
\end{equation}
Now it is useful to recall that in the single-particle formalism we would start with the transition amplitude ($t_0=0$):
\begin{equation}
\langle \vec{x},t\vert \vec{p}\rangle = \frac{1}{\sqrt{V}} e^{i\left( \vec{p}\cdot\vec{x}-t \sqrt{\vec{p}^2+m^2}\right)} \quad ,\label{eq:TransAmpl}
\end{equation}
which satisfies the Schr\"{o}dinger equation
\begin{align}
i\frac{\partial}{\partial t}\langle \vec{x},t\vert \vec{p}\rangle &= \langle \vec{x},t\vert H \vert \vec{p}\rangle\notag \\
&= \sqrt{\vec{p}^2+m^2}\langle \vec{x},t\vert \vec{p}\rangle\quad .\label{eq:SchroedingerEqnBraKetNotation}
\end{align}

One might wonder as to whether it is reasonable to define a probability amplitude $\langle \vec{x},t\vert \vec{x}_0\rangle$ also in relativistic quantum mechanics.
To find out let us begin with
\begin{align*}
\langle \vec{x},t\vert \vec{x}_0\rangle &= \sum_{\vec{p}} \langle \vec{x},t\vert \vec{p}\rangle\langle\vec{p}\vert\vec{x}_0\rangle,\qquad \left(\langle\vec{p}\vert\vec{x}_0\rangle = \frac{1}{\sqrt{V}}e^{-i\vec{x}_0\cdot \vec{p}} \right)\\
&= \sum_{\vec{p}} \frac{1}{V}e^{i\left(\vec{p}\cdot (\vec{x}-\vec{x}_0) - t\sqrt{\vec{p}^2+m^2}\right)},\quad \vec{r}:=\vec{x}-\vec{x}_0\\
&\overset{V\rightarrow\infty}{=} \int \frac{1}{(2\pi)^3} e^{i\left(\vec{p}\cdot\vec{r}-t\sqrt{\vec{p}^2+m^2} \right)} d^3\vec{p}\\
&= \frac{1}{(2\pi)^3}\int_0^\infty p^2 dp\ 2\pi\int_{-1}^{1} e^{iprz}e^{-it\sqrt{\vec{p}^2+m^2}} dz\\
&\overset{z=\cos\Theta}{=} -\frac{4\pi}{(2\pi)^3}\frac{1}{r}\frac{\partial}{\partial r}\frac{1}{2}\int_{-\infty}^{\infty} e^{ipr}e^{-it\sqrt{\vec{p}^2+m^2}}dp\quad .
\end{align*}
Here we change variables: $p=m\sinh\phi,\  \cosh\phi=\sqrt{1+\sinh^2\phi}$ to obtain:
\begin{align}
\langle\vec{x},t\vert \vec{x}_0\rangle &= -\frac{1}{(2\pi)^2}\frac{1}{r}\frac{\partial}{\partial r}\int m\cosh\phi\  e^{i\left(mr\sinh\phi - mt\cosh\phi\right)}d\phi\notag\\
&= -\frac{1}{(2\pi)^2}\frac{1}{r}\frac{\partial}{\partial r}i\frac{\partial}{\partial t}\int e^{i\left(mr\sinh\phi - mt\cosh\phi\right)}d\phi\quad .\label{eq:TransAmplParametric}
\end{align}

If we believe in causality we want $\langle \vec{x},t\vert\vec{x}_0\rangle$ to be zero for two points in a space-like relation:
\begin{equation*}
\left( \vec{x}-\vec{x}_0\right)^2 -t^2 = r^2-t^2 >0 ,\quad \text{space-like} (r>t)\quad .
\end{equation*}
So let us compute the integral \eqref{eq:TransAmplParametric} for $r>t$, and since a particle of mass $m> 0$ cannot travel with the speed $v \geq c$, we expect the integral to vanish.
For $r>t$:
\begin{equation*}
\left.\begin{aligned}
mr &= \lambda \cosh\phi_0 \\
mt &= \lambda \sinh\phi_0
\end{aligned}\right\rbrace\quad \lambda:=m\sqrt{r^2-t^2}\quad .
\end{equation*}
In \eqref{eq:TransAmplParametric} we then obtain:
\begin{align*}
mr\sinh\phi-mt\cosh\phi &= m\sqrt{r^2-t^2}\left(\sinh\phi\cosh\phi_0-\cosh\phi\sinh\phi_0\right)\\
&= m\sqrt{r^2-t^2}\sinh\left(\phi-\phi_0\right)\quad .\\
\intertext{Changing $\phi - \phi_0\rightarrow\phi$ we obtain for \eqref{eq:TransAmplParametric}: }
\int e^{im\sqrt{r^2-t^2}\sinh\phi}d\phi &= 2\int_0^\infty \cos\left( m\sqrt{r^2-t^2}\sinh\phi\right)d\phi\\
&\overset{\psi:=\sinh\phi}{=} 2\int_0^\infty \frac{\cos\left(m\sqrt{r^2-t^2}\psi\right)}{\sqrt{1+\psi^2}}d\psi\\
&= 2 K_0 (m\sqrt{r^2-t^2})\quad .
\end{align*}
This yields
\begin{equation}
\langle \vec{x},t\vert\vec{x}_0\rangle = -\frac{2i}{(2\pi)^2} \frac{1}{r}\frac{\partial}{\partial r}\frac{\partial}{\partial t}K_0(m\sqrt{r^2-t^2})\quad .\label{eq:TransAmplParametricSimplified}
\end{equation}

But $K_0(m\sqrt{r^2-t^2})$ and derivatives thereof are unequal zero for $r^2>t^2$, e.g. if $r\gg t$, then:
\begin{equation}
K_0(m\sqrt{r^2-t^2})\cong \sqrt{\frac{\pi}{2m\sqrt{r^2-t^2}}} e^{-m\sqrt{r^2-t^2}}\quad .\label{eq:Asymptotic_K0}
\end{equation}
Hence, the probability amplitude for finding the particle outside the light cone is non-zero - it drops exponentially. Hence it can be found in an area which is
forbidden by special relativity. We are in great trouble.

Now, let us use instead the language of second quantization.
We create a particle at $(\vec{x}_0,t_0=0)$ and detect it at $(\vec{x},t)$:
\begin{align*}
\psi^\dagger(\vec{x}_0) &= \sum_{\vec{p}'} \frac{1}{\sqrt{V}}a^\dagger(\vec{p}')e^{-i\vec{p}'\cdot \vec{x}_0}\\
\psi(\vec{x},t) &= \sum_{\vec{p}} \frac{1}{\sqrt{V}}a(\vec{p})e^{i(\vec{p}\cdot \vec{x}-t\sqrt{\vec{p}^2+m^2})}\quad .
\end{align*}
We wish to calculate the commutator, which is related to the probability amplitude of finding the particle at $(\vec{x},t)$ when it came into existence
at $\vec{x}_0$ at time $t_0=0$:
\begin{equation}
\left[ \psi(\vec{x},t),\psi^\dagger (\vec{x}_0)\right] = \sum_{\vec{p}} \frac{1}{V}e^{i\left(\vec{p}\cdot (\vec{x}-\vec{x}_0) - t\sqrt{\vec{p}^2+t^2}\right)}\quad .\label{eq:PsiFldCommutatorDiffTime}
\end{equation}
This looks exactly like the expression we had for $\langle \vec{x},t\vert\vec{x}_0\rangle$ in the single-particle description. Therefore the commutator will not vanish if the two points
$\vec{x},\vec{x}_0$ are in a space-like relation (c.f. Eq. \eqref{eq:TransAmplParametricSimplified}). Even worse: for equal times we obtain:
\begin{equation*}
\left[ \psi(\vec{x},t),\psi^\dagger (\vec{x}_0)\right]_{t=0} = \delta^3(\vec{x}-\vec{x}_0)\quad .
\end{equation*}
Not only have we violated causality - meaning no signal can travel faster than light. In addition, we find a breakdown of simultaneity, i.e., we cannot give the commutator an invariant
meaning outside the light cone. In order to remedy the whole situation, let us start anew by finding a relativistic invariant expression for the commutator.
The first step is to write
\begin{align*}
\left[ \psi(\vec{x},t),\psi^\dagger (\vec{x}_0)\right] &= \int \frac{1}{(2\pi)^3}e^{i\left(\vec{p}\cdot(\vec{x}-\vec{x}_0)-t\sqrt{\vec{p}^2+m^2}\right)}d^3\vec{p}\\
&= \int\frac{d^3\vec{p}}{(2\pi)^3}\int dp^0 e^{i\left(\vec{p}\cdot(\vec{x}-\vec{x}_0) - p^0t\right)}\Theta(p^0)\delta(p^0-\sqrt{\vec{p}^2+m^2})\\
= \int \frac{d^3\vec{p}}{(2\pi)^2}\int dp^0 e^{i\left(\vec{p}\cdot(\vec{x}-\vec{x}_0) - p^0t\right)} &\Theta(p^0) \delta\left[ (p^0-\sqrt{\vec{p}^2+m^2})(p^0+\sqrt{\vec{p}^2+m^2})\right]
2\sqrt{\vec{p}^2+m^2}
\end{align*}
or
\begin{align}
\left[ \psi(\vec{x},t),\psi^\dagger (\vec{x}_0)\right] &= \int\frac{d^3\vec{p}}{(2\pi)^3}\int dp^0 \Theta(p^0)2\sqrt{\vec{p}^2+m^2}\delta\left({p^0}^2-\vec{p}^2-m^2\right)
e^{i\left(\vec{p}\cdot(\vec{x}-\vec{x}_0) - p^0t\right)}\notag \\
&= \int\frac{1}{(2\pi)^3}\Theta(p^0) 2\sqrt{\vec{p}^2+m^2} \delta(p^2+m^2)e^{ip\cdot(x-x_0)} d^4p\quad .\label{eq:PsiFldCommutatorRelInv}
\end{align}
In our metric, $p^\mu =(\vec{p},p^0),p^2=\vec{p}^2-{p^0}^2,(x-x_0)=(\vec{x}-\vec{x}_0,t)$.

Without the factor $2\sqrt{\vec{p}^2+m^2}$ the integral in \eqref{eq:PsiFldCommutatorRelInv} is a Lorentz scalar, i.e., an invariant function under proper ortochronous Lorentz transformation:
\begin{equation*}
\int \frac{1}{(2\pi)^3}\Theta(p^0)\delta(p^2+m^2)e^{ip\cdot y} d^4p = F(y^2)\quad .
\end{equation*}
From our decomposition of the $\delta$-function,
\begin{equation*}
\Theta(p^0)\frac{1}{2\sqrt{\vec{p}^2+m^2}}\left[ \delta(p^0-\sqrt{\vec{p}^2+m^2}) + \delta(p^0+\sqrt{\vec{p}^2+m^2})\right] = \Theta(p^0)\delta(p^2+m^2)\quad ,
\end{equation*}
we see that the integral in \eqref{eq:PsiFldCommutatorRelInv} picks up a contribution from the top sheet of the mass shell hyperboloids.

Finally, to get rid of the factor $2\sqrt{\vec{p}^2+m^2}$ in \eqref{eq:PsiFldCommutatorRelInv}, we define a new operator:
\begin{equation}
\phi(\vec{x},t) := \sum_{\vec{p}} \frac{1}{\sqrt{V}}e^{i\left(\vec{p}\cdot\vec{x}-\omega_pt\right)}\frac{1}{\sqrt{2 \left(\vec{p}^2+m^2 \right)^{\scriptstyle \frac{1}{2} } } }a(\vec{p}),\quad \omega_p:=\sqrt{\vec{p}^2+m^2}\quad .\label{eq:PhiOpDef}
\end{equation}

The new commutator function is then given by
\begin{align}
\left[ \phi(\vec{x},t), \phi^\dagger (\vec{x}_0)\right] &= \int \frac{d^3\vec{p}}{(2\pi)^3}\frac{1}{2\sqrt{\vec{p}^2+m^2}} e^{i\left(\vec{p}\cdot(\vec{x}-\vec{x}_0)-t\sqrt{\vec{p}^2+m^2}\right)} \notag\\
&= \int\frac{d^4p}{(2\pi)^3}\Theta(p^0)\delta(p^2+m^2)e^{ip\cdot(x-x_0)}\quad ,\label{eq:PhiCommutatorR3,1}
\end{align}
which is an invariant function. Therefore $\phi$ and not $\psi$ is the appropriate operator.

For later purposes it is useful to calculate the $\int dp^0$ term in \eqref{eq:PhiCommutatorR3,1}, which reveals the invariant measure in momentum space:
\begin{equation}
d\omega_p = \frac{1}{2\omega_p}\frac{d^3\vec{p}}{(2\pi)^3},\quad \omega_p=\sqrt{\vec{p}^2+m^2}\quad .\label{eq:InvMeasureMomentumSpace}
\end{equation}
Having given $[\phi,\phi^\dagger]$ an invariant meaning we look at it in the frame where $t=0$:
\begin{align}
\left[\phi(\vec{x},t),\phi^\dagger(\vec{x}_0)\right]_{t=0} &= \int \frac{d^3\vec{p}}{(2\pi)^3} \frac{1}{2\sqrt{\vec{p}^2+m^2}}e^{i\vec{p}\cdot(\vec{x}-\vec{x}_0)}\notag\\
&= \int e^{i\vec{p}\cdot (\vec{x}-\vec{x}_0)}d\omega_p\quad ,\label{eq:PhiCommutatorSimultaneous}
\end{align}
which is not zero!

So our new commutator, although relativistic invariant, still violates causality (compare to the discussion of the transformation amplitude $\langle\vec{x},t\vert\vec{x}_0\rangle$).
Our next goal is therefore to restore causality.
By the way, we can use the operator \eqref{eq:PhiOpDef},
\begin{equation*}
\phi(\vec{x},t)=\frac{1}{\sqrt{V}}\sum_{\vec{p} } e^{i\left(\vec{p}\cdot\vec{x}-\omega_pt\right)} \frac{1}{2\sqrt{\omega_p}}a(\vec{p})
\end{equation*}
to build a wave packet,
\begin{equation}
\langle 0\vert \phi(\vec{x},t) = \sum_{\vec{p}} \frac{1}{\sqrt{V}}e^{i\left(\vec{p}\cdot\vec{x}-\omega_pt\right)} \frac{1}{2\sqrt{\omega_p}}\langle \vec{p}\vert\quad 
\bigl(\langle 0\vert a(\vec{p})=\langle\vec{p}\ \vert\bigr)\quad .\label{eq:PhiWavePacket}
\end{equation}
The inner product of two wave packets is
\begin{align}
\langle 0\vert \phi(\vec{x},t)\phi^\dagger(\vec{x}_0)\vert 0\rangle &= \langle 0 \vert \left[ \phi(\vec{x},t),\phi^\dagger(\vec{x}_0)\right]\vert0\rangle\notag\\
&= \int \frac{d^4p}{(2\pi)^3}\Theta(p^0)\delta(p^2+m^2)e^{ip\cdot(x-x_0)}\quad ,\label{eq:PhiWavePacketsIP}
\end{align}
which means that the localization of a particle has an invariant meaning, i.e., again, looks the same for all observers.

However, the description is still complicated from the point of causality.
Although $[\phi,\phi^\dagger]$ is Lorentz invariant, it does not vanish at $t=0$:
\begin{equation}
\left[\phi(\vec{x},t),\phi^\dagger(\vec{x}_0)\right] = \int \frac{d^3\vec{p}}{(2\pi)^3} \frac{1}{2\omega_p}e^{i\vec{p}\cdot (\vec{x}-\vec{x}_0)}e^{-i\omega_p t}\quad .\label{eq:PhiCommutatorR3}
\end{equation}
Recall that this expression was obtained by using $\phi,\phi^\dagger$ given above and the commutation relation $[a,a^\dagger]=1$.

In order to construct a commutator that vanishes for $t=0$ we have to subtract something from our former expression \eqref{eq:PhiCommutatorR3}, namely the second term in
\begin{equation}
\int \frac{d^2\vec{p}}{(2\pi)^3}\frac{1}{2\omega_p}\left( e^{i\vec{p}\cdot(\vec{x}-\vec{x}_0)} e^{-i\omega_pt} - e^{-i\vec{p}\cdot(\vec{x}-\vec{x}_0)} e^{i\omega_pt}\right)\quad .\label{eq:PhiCancellationExpression}
\end{equation}
For the second term we use the lower sheet of the hyperboloid, i.e., take the solution $p^0 = -\sqrt{\vec{p}^2+m^2}$. Then $\Theta(-p^0)\delta(p^2+m^2)$ selects the bottom sheet as
$\Theta(p^0)\delta(p^2+m^2)$ picks out the top sheet - our situation so far.

Now comes the point: the minus sign in \eqref{eq:PhiCancellationExpression} plays a significant role. It is the same minus sign that occurs if we look at $[a,a^\dagger]=1$ and write instead
$[a^\dagger,a]=-1$.
Therefore, if we interchange the role of creation and destruction operators, we can convert the minus sign into a plus sign:
\begin{equation}
\tilde{\phi}(\vec{x},t)=\frac{1}{\sqrt{V}}\sum_{\vec{p}} \frac{1}{\sqrt{2\omega_p}} \left( e^{i(\vec{p}\cdot\vec{x}-\omega_pt)}a(\vec{p})+e^{-i(\vec{p}\cdot\vec{x}-\omega_pt)}b^\dagger(\vec{p})\right)\quad ,\label{eq:PhiOpDefMultiOp}
\end{equation}
with $[a,a^\dagger]=1,[a,b]=0,[b,b^\dagger]=1,\omega_p=+\sqrt{\vec{p}^2+m^2}$.
Then, with this new object $\tilde{\phi}\rightarrow\phi$ we obtain:
\begin{equation*}
\left[\phi(\vec{x},t),\phi^\dagger(\vec{x}_0)\right] = \int \frac{d^2\vec{p}}{(2\pi)^3}\left( e^{i\vec{p}\cdot(\vec{x}-\vec{x}_0)} e^{-i\omega_pt} -
e^{-i\vec{p}\cdot(\vec{x}-\vec{x}_0)} e^{i\omega_pt}\right)\quad ,
\end{equation*}
which yields at last $[\phi(\vec{x},t),\phi^\dagger(\vec{x}_0)]_{t=0} = 0$ for $\vec{x}\neq \vec{x}_0$ and $\vec{x} = \vec{x}_0$.
The invariant form of $[\phi,\phi^\dagger]$ becomes obvious when we write ($y:=(\vec{x}-\vec{x}_0,t)$):
\begin{align}
\left[\phi(\vec{x},t),\phi^\dagger(\vec{x}_0)\right] &= \int \frac{d^3pdp^0}{(2\pi)^3}\left(\Theta(p^0)\delta(p^2+m^2)e^{ip\cdot y}-\Theta(-p^0)\delta(p^2+m^2)e^{ip\cdot y}\right)\notag\\
&=: i\Delta(y)\quad .\label{eq:PJCFDef}
\end{align}
This so-called Pauli-Jordan invariant commutator function is zero for $t=0$ and invariant for space-like distances $y^2>0$ with $y^2=(\vec{x}-\vec{x}_0)^2-t^2$.
Using 
\begin{equation}
\epsilon(p^0) = \left\lbrace \begin{array}{lr}
+1 &,\quad p^0>0\\
-1 &,\quad p^0<0
\end{array}\right\rbrace =\Theta(p^0)- \Theta(-p^0)\quad ,
\end{equation}
we obtain the final form
\begin{equation}
\left[\phi(\vec{x},t),\phi^\dagger(\vec{x}_0,0)\right] = i\Delta(y) = \int \frac{d^4p}{(2\pi)^3}\epsilon(p^0)\delta(p^2+m^2)e^{ip\cdot y}\quad .\label{eq:PhiCommutatorFinalForm}
\end{equation}
If we calculate $\langle 0\vert \phi(\vec{x},t)\phi^\dagger(\vec{x}_0,0)\vert 0\rangle$ and insert the expressions for our old $\tilde{\phi},\tilde{\phi}^\dagger$ from \eqref{eq:PhiOpDefMultiOp},
the result becomes identical to our old calculation with the $\phi$ from \eqref{eq:PhiWavePacketsIP}. Therefore the vacuum expectation value is the same - since we assume that the 
particles associated with the operators $a$ and $b$ have the same mass.

If we take $b= a$, then
\begin{equation}
\phi(\vec{x},t)=\frac{1}{\sqrt{V}}\sum_{\vec{p}} \frac{1}{2\sqrt{\omega_p}} \left( e^{i(\vec{p}\cdot\vec{x}-\omega_pt)}a(\vec{p}) + e^{-i(\vec{p}\cdot\vec{x}-\omega_pt)}a^\dagger(\vec{p})\right)
\quad ,\label{eq:PhiOpDefMultiOpSameType}
\end{equation}
meaning $\phi^\dagger=\phi$, i.e., $\phi$ is Hermitean.

If we consider \eqref{eq:PhiOpDefMultiOp} again and calculate the derivative $i\tfrac{\partial\phi}{\partial t}$, we find that it does not satisfy a first-order differential equation but
a second-order one:
\begin{equation}
\frac{\partial^2}{\partial t^2}\phi(\vec{x},t) = \left(\vec{\nabla}^2-m^2\right) \phi(\vec{x},t)\quad ,\label{eq:KGDiffEq2ndOrder}
\end{equation}
which is local in space-time; $a$ and $b^\dagger$ correspond to the two constants of integration.
The equation
\begin{equation}
\left(\frac{\partial}{\partial t}\right)^2\phi = \left(\vec{\nabla}^2-m^2\right)\phi\  ,\label{eq:KGFSEqn}
\end{equation}
is called the Klein-Gordon-Fock-Schr\"{o}dinger (K.G.F.Sch.) equation. It is a local relativistic equation.

%%remark
\textbf{Remark} When we constructed the local operator $\phi$ we used $[a,a^\dagger]=1$. At one point it was necessary - in order to restore causality - to use the minus sign in 
$[b^\dagger, b]=-1$.
If we now would go back and use instead an anti-commutation relation $aa^\dagger+ a^\dagger a=\lbrace a,a^\dagger\rbrace=1$, our procedure to construct a local field operator would fail.
The particles have to have spin.

Conventional textbooks start with the K.G.F.Sch. equation
\begin{equation}
\left( -\partial^2+m^2\right)\phi(\vec{x},t) = \left[ \left(\frac{\partial}{\partial t}\right)^2 - \vec{\nabla}^2 + m^2\right] \phi(\vec{x},t) = 0\label{eq:KGFSEqnR31}
\end{equation}
and look for solutions with the two constants of integration specified by $\phi(\vec{x},0)$ and $\dot{\phi}(\vec{x},0)$ and find that \eqref{eq:PhiOpDefMultiOp}
is the local solution to the K.G.F.Sch. equation. This is the path that mathematicians would take. Our procedure is more suited to physics-minded students and teachers.

We found in \eqref{eq:PhiCommutatorFinalForm} a most important invariant function $\Delta$ (Pauli-Jordan). Let us write it as
\begin{equation*}
\Delta(x,\kappa_0^2) = -i (2\pi)^{-3} \int (dk) e^{ikx} \epsilon(k^0)\delta(k^2+\kappa_0^2),
\end{equation*}
where $(dk)=dk_0dk_1dk_2dk_3$.
But there is another invariant function:
\begin{equation*}
\Delta^{(1)}(x,\kappa_0^2) = (2\pi)^{-3} \int (dk) e^{ikx}\delta(k^2+\kappa_0^2)\quad .
\end{equation*}
Evidently, both are solutions to the K.G. equation:
\begin{equation*}
(\kappa_0^2-\partial^2 ) \lbrace \Delta, \Delta^{(1)}\rbrace = 0\quad .
\end{equation*}
The two functions fulfill different boundary conditions and  different symmetry properties. In fact,
\begin{align*}
\Delta(-x) &= -\Delta(x),\quad\text{since }\epsilon\text{ is odd}\\
\Delta^{(1)} (-x) &=\Delta^{(1)} (x)\quad .
\end{align*}
They both share invariance under proper ortochronous Lorentz transformation:
\begin{equation*}
 \left( \Delta, \Delta^{(1)}\right)(\Lambda x) =  \left( \Delta, \Delta^{(1)}\right) (x)\quad .
\end{equation*}

The basic fact about $\Delta$ is that it is zero outside the light cone while $\Delta^{(1)}$ reaches into the space-like sector where it dies out exponentially.
In other words, microcausality is realized by $\Delta$ and not determined by $\Delta^{(1)}$! It is due to the behavior of $\Delta$, i.e., the disappearance of the commutator of the
fields, that measuring the field at $x_0$ can have no consequence on measuring the field at $x$ since the points are not causally connected.
However both $\Delta$ and $\Delta^{(1)}$ are the basic functions for constructing the remaining invariant functions. More about their explicit expression in coordinate space will be
given and thoroughly discussed in the remaining chapters.

%%%%%%%%%%%%%%%%%%%%%%%%%%%%%%%%%%%%%%%%%%%%%%%%%%%%%%%%%%%
%%%%%%%%%%%%%%%%%%%%%%%%%%%%%%%%%%%%%%%%%%%%%%%%%%%%%%%%%%%%%%%%%%%%%%%%%%%%%%%%%%%%%%%%%%%%%%%%%%%%%%%%%%%
%%%%%%%%%%%%%%%%%%%%%%%%%%%%%%%%%%%%%%%%%%%%%%%%%%%%%%%%%%%

\section{Selection of invariant commutation and propagation functions}
In the last chapter, the Pauli-Jordan commutation function was constructed, starting from a scalar field:
\begin{equation}
\Delta(x) = -i\frac{1}{(2\pi)^3}\int e^{ik\cdot x}\epsilon(k_0)\delta(k^2+\mu^2)(dk)\quad .\label{eq:PJCFDef2.1}
\end{equation}
We also mentioned a second invariant function:
\begin{equation}
\Delta^{(1)}(x) = \frac{1}{(2\pi)^3}\int e^{ik\cdot x}\delta(k^2+\mu^2)(dk)\quad .
\end{equation}
Although they both satisfy the K.G.F. equation, they play a totally different role in scalar quantum field theory (Q.F.T.). While $\Delta(x)$ vanishes if
$x^2>0$ (space-like argument), $\Delta^{(1)}(x)$ does not vanish for space-like distances(c.f. Appendix \ref{sec:Appendix} ). Instead, $\Delta^{(1)}$ extends into the space-like region,
dropping off on the scale of the Compton wavelength $\tfrac{1}{\mu}$. However, it is also the famous Feynman propagator function that reaches into the space-like region.

In our convention, $\Delta_F (x) = \Delta^{(1)}(x)+i\epsilon(x)\Delta(x),\  \Delta_F =:2i\Delta_c$.
Note that the inhomogeneous $\Delta_F$ is constructed from the two homogeneous invariant functions $\Delta$ and $\Delta^{(1)}$. For our purposes we will use
the momentum representation of $\Delta_c(x)$:
\begin{equation}
\Delta_c(x) = \frac{1}{(2\pi)^4}\int \frac{e^{ik\cdot x}}{k^2+\mu^2-i\epsilon} (dk)\quad .\label{eq:DeltaCMomentumRepresentation}
\end{equation}
Employing
\begin{equation*}
\frac{1}{k^2+\mu^2-i\epsilon} = i\int_0^\infty e^{-is(k^2+\mu^2-i\epsilon}ds\quad ,
\end{equation*}
we obtain:
\begin{align}
\Delta_c(x) &= \int\frac{(dk)}{(2\pi)^4} i\int_0^\infty e^{-is(k^2+\mu^2)} e^{ik\cdot x}ds\notag\\
&= \frac{i}{(2\pi)^4}\int_0^\infty ds\  e^{-is\mu^2} \int e^{-isk^2+ik\cdot x}(dk)\quad .\label{eq:DeltaCMomentumHilbertTransform}
\end{align}
The $k$-integral in the previous equation is given by
\begin{equation}
\int e^{-isk^2+ik\cdot x}(dk) = -i\frac{\pi^2}{s^2} e^{i\frac{x^2}{4s}}\quad .
\end{equation}
Therefore
\begin{align*}
\Delta_c(x) &= \frac{1}{(2\pi)^4}\int_0^\infty e^{-is\mu^2}\frac{\pi^2}{s^2}e^{i\frac{x^2}{4s}} ds = \frac{1}{16\pi^2}\int_0^\infty \frac{1}{s^2} e^{-is\mu^2}e^{i\frac{x^2}{4s}}ds\\
\Delta_c(x) &: \begin{cases}
\sim \text{Hankel function}\sim\frac{1}{\sqrt{-x^2}},\text{propagation},x^2<0 & \text{inside light cone}\\
\sim K_1\text{ function}\sim e^{-\text{const.}\sqrt{x^2}},\text{not propagation}, x^2>0 & \text{outside light cone .}
\end{cases}
\end{align*}
Here is the result for the causal Green's function $\Delta_c(x)$:
\begin{align*}
\Delta_c(x) &= \frac{1}{4\pi}\delta(x^2)+\Theta(x^2)\frac{i\mu}{4\pi^2\sqrt{x^2}} K_1(\mu\sqrt{x^2}) - \Theta(-x^2)\frac{\mu}{8\pi\sqrt{-x^2}} H_1^{(2)}(\mu\sqrt{-x^2})\\
&= \frac{1}{4\pi}\delta(x^2) - \Theta(-x^2)\frac{\mu}{8\pi\sqrt{-x^2}}\left[ J_1(\mu\sqrt{-x^2}) -i N_1(\mu\sqrt{-x^2})\right]\\
&+\Theta(x^2)\frac{i\mu}{4\pi\sqrt{x^2}}K_1(\mu\sqrt{x^2})\quad .
\end{align*}
For the Pauli-Jordan function \eqref{eq:PJCFDef2.1} we use, from the detailed calculations in the appendix ($\Delta(x) := -2\bar{\Delta}(x)\epsilon(x)$), the result 
\eqref{eq:AppDeltaBarCompact}:
\begin{align*}
\bar{\Delta}(x) &= \frac{1}{4\pi}\delta(x^2) -\frac{\mu^2}{8\pi}\mathfrak{Re}\left[\frac{H_1^{(1)}(\mu\sqrt{-x^2}) }{\mu\sqrt{-x^2}}\right]\\
\mathfrak{Re}\left[\frac{H_1^{(1)}(\mu \sqrt{-x^2} ) }{\mu\sqrt{-x^2}}\right] &= \begin{cases}
\frac{J_1(\mu \sqrt{-x^2} )}{\mu \sqrt{-x^2}} & x^2 <0\\
0 & x^2 >0\quad .
\end{cases}
\end{align*}
$\bar{\Delta}(x)$ and therefore $\Delta(x)$ vanishes if $x^2>0$ (space-like distance), which is how we constructed the Pauli-Jordan commutator function. Finally,
\begin{equation*}
\Delta(x) = -\frac{1}{2\pi}\epsilon(x_0)\delta(x^2)+\frac{\mu}{4\pi\sqrt{-x^2}}\epsilon(x_0)\Theta(-x^2)J_1(\mu\sqrt{-x^2})\quad .
\end{equation*}
Unlike the situation for $\bar{\Delta}(x)$ there is no discontinuity at $x^2=0$ for $\Delta^{(1)}(x)$ and therefore $\Delta^{(1)}$ does not vanish for space-like distances($x^2>0$).
To show this we refer to the appendix:
\begin{align*}
\Delta^{(1)}(x) &= \frac{1}{\pi}\PV\int\Delta(x-\epsilon\tau)\frac{1}{\tau}d\tau\quad .
\intertext{Here we insert}
\Delta(x) &= -\frac{i}{(2\pi)^3}\int e^{ik\cdot x} \delta(k^2+\mu^2)\epsilon(k_0)(dk)\quad ,
\intertext{such that}
\Delta^{(1)}(x) &= -\frac{i}{\pi}\frac{1}{(2\pi)^3}\int(dk)\epsilon(k)
\underset{{i\pi\epsilon(k)}}{\underbrace{\left[\PV\int\frac{1}{\tau}e^{-ik\cdot\epsilon\tau}d\tau\right] }} e^{ik\cdot x}\delta(k^2+\mu^2)\quad ,
\intertext{which yields:}
\Delta^{(1)}(x) &= \frac{1}{(2\pi)^3}\int e^{ik\cdot x}\delta(k^2+\mu^2)(dk)\quad ,
\end{align*}
which is nothing but the momentum representation of $\Delta^{(1)}(x)$.
In \eqref{eq:AppDelta1FinalRepresentaion} we present the final result for $\Delta^{(1)}(x)$:
\begin{equation}
\Delta^{(1)}(x) = \frac{\mu}{4\pi}\frac{1}{\sqrt{-x^2}}\Theta(-x^2)N_1(\mu\sqrt{-x^2})+\frac{\mu}{2\pi^2}\frac{1}{\sqrt{x^2}}\Theta(x^2)K_1(\mu\sqrt{x^2})\quad .\label{eq:Delta1FinalResultShortForm}
\end{equation}
For a space-like distance and equal time we have $(x-x')\rightarrow (\vec{x}-\vec{x}')=\vec{z},|\vec{z}|=:r$. Then the last term in \eqref{eq:Delta1FinalResultShortForm} allows us to study
its behavior for large $r$:
\begin{align*}
\Delta^{(1)}(\vec{z},0) &= \frac{\mu}{2\pi^2 r}K_1(\mu r) = 2\frac{\mu}{4\pi^2r}K_1(\mu r)\\
&\overset{r\rightarrow\infty}{\sim} \frac{2\sqrt{\mu}}{(2\pi r)^{\frac{3}{2}} }e^{-\mu r}\\
&\overset{x'=0}{=} \frac{2\sqrt{\mu} }{ \left(2\pi\sqrt{x^2}\right)^{ \frac{3}{2} } } e^{-\mu\sqrt{x^2} },\quad x^2>0\quad ,
\end{align*}
which means that $\Delta^{(1)}$ drops as $e^{-\mu r}$ for $\mu\sqrt{x^2}\gg 1$, where $\tfrac{1}{\mu}\sim$ Compton wavelength.

%%%%%%%%%%%%%%%%%%%%%%%%%%%%%%%%%%%%%%%%%%%%%%%%%%%%%%%%%%%
%%%%%%%%%%%%%%%%%%%%%%%%%%%%%%%%%%%%%%%%%%%%%%%%%%%%%%%%%%%%%%%%%%%%%%%%%%%%%%%%%%%%%%%%%%%%%%%%%%%%%%%%%%%
%%%%%%%%%%%%%%%%%%%%%%%%%%%%%%%%%%%%%%%%%%%%%%%%%%%%%%%%%%%

\section{Pair Production of Charged Scalar Particles}
In the last chapter of their paper\cite{Pauli1934} Pauli and Weisskopf calculate the probability for a scalar charged pair production with photons in the presence of a Coulomb
field. The theory of particle production ($e^+,e^-$) by $\gamma$ rays in the Coulomb field of a nucleus had already been calculated by Bethe and Heitler with the aid of 
Dirac's hole theory\cite{Bethe1934}. According to Pauli: "The most interesting part of our theory is that the energy (of the produced particles) is always positive automatically (namely,
without using a superfluous hypothesis such as hole theory)."

We will refrain from calculating any processes of scalar QED. The textbook literature offers abundant examples, including scattering as well as bound-state problems. We will instead finish
this article by working out the problem of pair production in scalar QED in presence of an external, constant, electromagnetic field. Research in producing spinor as well as 
scalar charged particles in external fields is, at the moment, of worldwide interest.

Let us define our problem more explicitly. It is known that the Lagrangian of a free electromagnetic field is given by $\mathcal{L}_0 = -\tfrac{1}{4} F_{\mu\nu}^2(x)$.
We want to find out how this Lagrangian is modified if the quantum vacuum is taken into account. More specifically, we want to compute a charged scalar particle loop to all orders in a constant,
prescribed, external field and determine how it affects the free electromagnetic Lagrangian.
In other words, we want to calculate the effective Lagrangian for scalar QED. I will present as many details as possible. But for a thorough understanding of the whole problem,
scalar as well as spinor particles in the loop, I would like to invite the reader to consult the Springer Lecture Notes \cite{Dittrich1985} or the monography \cite{Dittrich2000}.

Here are some basic facts:
\begin{itemize}
\item Hamiltonian: $H=\Pi^2$, $E$ field and $H$ field in $x_3$ direction
\item Gauge, background field $A_\mu(x) = \left( -\frac{1}{2}Hx_2,\frac{1}{2}Hx_1, -Ex^0;0\right)$
\item Non-zero field components $F_{12} = -F_{21} = H, F_3^0 = E$
\end{itemize}
\begin{equation}
\Pi_\mu = p_\mu - eA_\mu : \begin{cases}
\left. \begin{array}{c}
\Pi_1 = p_1+\frac{e}{2}Hx_2\\
\Pi_2 = p_2 - \frac{e}{2}Hx_1
\end{array} \right\rbrace & \left[\Pi_1,\Pi_2\right] = ieH\\
\Pi_3 = p_3 +eEx^0 & \\
\Pi^0 = p^0 &
\end{cases}\quad .
\end{equation}
In general $[\Pi_\mu,\Pi_\nu] = ieF_{\mu\nu}$.
\begin{align*}
\Pi^2 = (p-eA)^2 &= \left(\Pi_1^2+\Pi_2^2\right)+\left(\Pi_3^2-\Pi_0^2\right) = \Pi_\perp^2 + \Pi_\parallel^2,\quad [\Pi_\perp,\Pi_\parallel] = 0\\
&= \left(p_1+\frac{e}{2} Hx_2\right)^2+\left(p_2-\frac{e}{2} Hx_1\right)^2 + \left(p_3+eEx^0\right)^2-{p^0}^2\  .
\end{align*}
Now, without derivation - it can be found in our Lecture Notes\cite{Dittrich1985} - the effective Lagrangian for a complex scalar K.G.F. field is given by
\begin{equation}
\mathcal{L}_{eff} = \mathcal{L}_0 + \mathcal{L}_0^{(1)}\quad ,
\end{equation}
where $\mathcal{L}_0^{(1)}$ denotes the one-loop (effective) Lagrangian
\begin{equation}
\mathcal{L}_0^{(1)} = -i\int_0^\infty \frac{1}{s}e^{-im^2s}\langle x \vert e^{-is\Pi^2}\vert x\rangle ds\quad .
\end{equation}
Hence we need to work out the following diagonal element($ x^0=t$):
\begin{equation*}
\langle x \vert e^{-is\Pi^2}\vert x\rangle = \langle x \vert e^{-is\Pi^2_\perp} e^{-is\Pi_\parallel^2}\vert x\rangle = 
\langle x_1 x_2 \vert e^{-is\Pi^2_\perp}\vert x_1x_2\rangle\langle x_3 x^0\vert e^{-is\Pi_\parallel^2}\vert x_3x^0 \rangle\quad .
\end{equation*}
The two transition amplitudes we need to calculate are
\begin{align*}
\int \langle x_1x_2\vert p_1p_2\rangle\langle p_1p_2\vert &e^{-is\left(\Pi_1^2+\Pi_2^2\right)} \vert p_1'p_2'\rangle\langle p_1'p_2'\vert x_1x_2\rangle dp_1dp_2dp_1'dp_2'\\
\int \langle x_3t\vert p_3\omega\rangle\langle p_3\omega\vert &e^{-is\left(\Pi_3^2-\Pi_0^2\right)} \vert p_3'\omega'\rangle\langle p_3'\omega'\vert x_3t\rangle dp_3d\omega dp_3'd\omega'\quad .
\end{align*}
Here we have to insert
\begin{align*}
\langle x_1x_2\vert p_1p_2\rangle &= \frac{1}{2\pi}e^{i(p_1x_1+p_2x_2)} & \langle p_1'p_2'\vert x_1x_2\rangle &= \frac{1}{2\pi}e^{-i(p_1'x_1+p_2'x_2)}\\
\langle x_3t\vert p_3\omega\rangle &= \frac{1}{2\pi}e^{i(p_3x_3-\omega t)} & \langle p_3'\omega'\vert x_3 t\rangle &= \frac{1}{2\pi}e^{-i(p_3'x_3-\omega't)}\quad .
\end{align*}
With a few steps in between, the result for the $E$ part is
\begin{align*}
\frac{1}{(2\pi)^2}\int e^{i(p_3-p_3')x_3}e^{-i(\omega-\omega')t} \langle p_3\omega\vert &e^{-is\left[ (eE)^2(x^0-\frac{p_3}{eE})^2-{p^0}^2\right]}\vert p_3'\omega'\rangle dp_3dp_3'd\omega d\omega'\\
&= \frac{1}{(2\pi)^2}\frac{\pi}{s} \frac{esE}{\sinh(esE)} = \frac{1}{4\pi s} \frac{esE}{\sinh(esE)}\\
&= \langle x_3x^0\vert e^{-is\Pi_\parallel^2}\vert x_3x^0\rangle\quad .
\end{align*}
For the $H$ term we take from elementary quantum mechanics the spectrum for $H=\left(\vec{p}-e\vec{A}\right)_\perp^2 : E_n =(2n+1)eH$ so that
(the factor $\tfrac{eH}{2\pi}$ takes into account the degeneracy per unit area of the Landau levels):
\begin{align*}
\langle x_1x_2\vert e^{-is\Pi_\perp^2} \vert x_1x_2\rangle &= \frac{eH}{2\pi} \sum_n e^{-(2n+1)(ieHs)} = \frac{1}{4\pi}\frac{eH}{\sinh(ieHs)}\\
&= \frac{1}{4\pi s}\frac{eHs}{i\sin(eHs)}\quad .
\end{align*}
Hence we obtain the diagonal element:
\begin{equation*}
\langle x \vert e^{-is\Pi^2}\vert x\rangle = \left( \frac{1}{4\pi s i}\frac{eHs}{\sin(eHs)}\right)\left( \frac{1}{4\pi s}\frac{eEs}{\sinh(eEs)}\right) = 
\frac{1}{16\pi^2}\frac{1}{is^2}\left[ \frac{eHs}{\sin(eHs)}\frac{eEs}{\sinh(eEs)}\right]
\end{equation*}
(c.f. also (5.15) in \cite{Dittrich1985}).

With this result we can write:
\begin{equation}
i\mathcal{L}_0^{(1)} = \frac{1}{16\pi^2}\frac{1}{i}\int_0^\infty \frac{1}{s^3}e^{-im^2s} \frac{eHs}{\sin(eHs)}\frac{eEs}{\sinh(eEs)}ds\quad .
\end{equation}
Our final result, with the necessary subtraction terms to produce a finite answer, is given by the Heisenberg-Euler(H.E.) Lagrangian for scalar electrodynamics $\mathcal{L}_{HE} := 
\mathcal{L}_0 + \mathcal{L}_0^{(1)}$ with
\begin{equation}
\mathcal{L}_0^{(1)}(H,E) = \frac{1}{16\pi^2}\int_0^\infty \frac{ds}{s^3}e^{-m^2s}\left[ \frac{eHs}{\sinh(eHs)}\frac{eEs}{\sin(eEs)} - 1 - \frac{e^2s^2}{6}(E^2-H^2)\right]\ .
\end{equation}
Let us limit ourselves to the case of a pure applied constant $H$ field. Then there are a number of ways to explicitly compute the $s$-integral, e.g., with dimensional regularization, $\zeta$-
function regularization, etc. They all agree with the answer
\begin{equation}
\mathcal{L}_0^{(1)}(H) = \frac{1}{64\pi^2}\left\lbrace \left[ 2m^4-\frac{2}{3}\left(eH\right)^2\right]\left(1+\ln\frac{m^2}{2eH}\right) -3m^4-(4eH)^2\zeta'\left(-1,\frac{m^2+eH}{2eH}\right)\right\rbrace .
\end{equation}
The conversion from a pure magnetic field to a pure electric field takes place by substitution $B\rightarrow -iE$. Like in spinor electrodynamics it is possible to compute
via $2\mathfrak{Im}\left[\mathcal{L}_0^{(1)}(E)\right]$ the pair creation probability for spin 0 charged scalar particles in an external constant $E$ field\cite{Dittrich2014}.
Here is the final formula:
\begin{equation}
2 \mathfrak{Im}\left[\mathcal{L}^{(1)}(E)\right] = (2s+1)\frac{(eE)^2}{(2\pi)^3}\sum_{n=1}^\infty \frac{(\pm 1)^{n+1}}{n^2}e^{-n\frac{\pi m^2}{eE}}\quad ,
\end{equation}
where the upper and lower signs correspond to $s=\tfrac{1}{2}$ and $s=0$, respectively.

It should be emphasized that the H.E. Lagrangian yields a totally non-perturbative theory for low-energy photons. It encodes a tremendous amount of physics,
like photon-photon scattering, vacuum polarization, determination of the $\beta$ function in scalar as well as spinor QED, etc. All together the Pauli-Weisskopf theory
is as far-reaching for spinless particles as the Dirac theory is for spin-$1/2$ particles - without ever mentioning Dirac's hole theory.
%%%%%%%%%%%%%%%%%%%%%%%%%%%%%%%%%%%%%%%%%%%%%%%%%%%%%%%%%%%
%%%%%%%%%%%%%%%%%%%%%%%%%%%%%%%%%%%%%%%%%%%%%%%%%%%%%%%%%%%%%%%%%%%%%%%%%%%%%%%%%%%%%%%%%%%%%%%%%%%%%%%%%%%
%%%%%%%%%%%%%%%%%%%%%%%%%%%%%%%%%%%%%%%%%%%%%%%%%%%%%%%%%%%

\appendix

\section{The Invariant $\Delta$ Functions\cite{Schwinger1949} }\label{sec:Appendix}

Since all the $\Delta$ functions can be expressed in terms of the two independent basic functions $\Delta(x)$ and $\Delta^{(1)}(x)$, we begin with the construction of the invariant function $\bar{\Delta}(x)$, which has a simple connection to the Pauli-Jordan function $\Delta(x)$
according to
\begin{align}
\bar{\Delta}(x) = -\frac{1}{2}\Delta(x)\epsilon(x) &= \frac{1}{2}\Delta(x)\frac{\epsilon_\mu x_\mu}{|\epsilon_\mu x_\mu|}\label{eq:AppDeltaBarDeltaRelation}\quad ,\\
\intertext{where}
\epsilon(x) = \frac{-(\vec{\epsilon}\cdot\vec{x}-\epsilon_0x_0 )}{|\epsilon_\mu x_\mu|} &= -\frac{\epsilon_\mu x_\mu}{|\epsilon_\mu x_\mu|} =\begin{cases}
1 & x_0>0\\
-1 & x_0 <0\quad .
\end{cases}\label{eq:AppDefinitionEpsilonR3,1}
\end{align}
This sign factor is effectively an invariant since only time-like vectors $x_\mu$, i.e., $x_0>|\vec{x}|$, are considered in \eqref{eq:AppDeltaBarDeltaRelation}.
$\epsilon_\mu$ is an arbitrary time-like vector with $\epsilon_0>0$. $\bar{\Delta}(x)$ satisfies the following equation:
\begin{equation}
\left( \partial^2-\kappa_0^2\right) \bar{\Delta}(x) = 0,\quad x_\mu \neq 0\quad .\label{eq:AppDeltaBarDiffEq}
\end{equation}
To evaluate the left side of this equation at the origin we consider
\begin{equation*}
\Lim\int_{\delta\omega} \left( \partial^2-\kappa_0^2\right) \bar{\Delta}(x)d\omega = \Lim\left[\int_{\sigma_+}\frac{\partial\bar{\Delta}(x)}{\partial x_\mu} d\sigma_\mu 
-\int_{\sigma_-}\frac{\partial\bar{\Delta}(x)}{\partial x_\mu} d\sigma_\mu \right],
\end{equation*}
in which the region of integration $\delta\omega := dx_0dx_1dx_2dx_3$ is extended between two space-like surfaces $\sigma_+$ and $\sigma_-$, which lie in the future and past, respectively,
relative to the origin, and coincide in the limit with the space-like surface $\sigma$ through the origin. Thus
\begin{align*}
\Lim\int_{\delta\omega} \left( \partial^2-\kappa_0^2\right) \bar{\Delta}(x)d\omega &= \Lim\left[ \int_{\sigma_+} \frac{\partial}{\partial x_\mu} \left(-\frac{1}{2}
\Delta(x)\epsilon(x)\right) d\sigma_\mu - \int_{\sigma_-} \frac{\partial}{\partial x_\mu} \left(-\frac{1}{2}\Delta(x)\epsilon(x)\right) d\sigma_\mu\right]\\
&= -\int_\sigma \frac{\partial}{\partial x_\mu}\Delta(x)d\sigma_\mu\  ,\  \text{for }\epsilon(x)=\begin{cases}1,&\quad x_0>0\\
-1,&\quad x_0<0\  .
\end{cases}
\end{align*}
Let us evaluate the integral on a flat space surface, i.e., $d\sigma_\mu = -id^3x$ and remember:
$\frac{\partial}{\partial x_0}\Delta(x) = -\delta(x)$. Then
\begin{align}
\Lim\int_{\delta\omega} \left( \partial^2-\kappa_0^2\right) \bar{\Delta}(x)d\omega &= -\frac{1}{i}\int(-\delta(x))(-id^3x)\quad .\label{eq:AppLimitSimplification}
\intertext{All this shows that}
\left( \partial^2-\kappa_0^2\right) \bar{\Delta}(x) &= -\delta(x)\quad\left( = -\delta(x_0)\delta(x_1)\delta(x_2)\delta(x_3)\right)\  .\label{eq:AppDeltaBarGreensFunction}
\end{align}
Evidently $\bar{\Delta}(x)$ plays the role of a four-dimensional Green's function. In terms of the integral representation:
\begin{equation}
\delta(x) = \frac{1}{(2\pi)^4}\int e^{ik_\mu x_\mu}(dk)\quad ,\label{eq:AppDiracDeltaItegralR4}
\end{equation}
where $(dk):=dk_0dk_1dk_2dk_3$,
we obtain as a particular solution of \eqref{eq:AppDeltaBarGreensFunction}:
\begin{align}
\bar{\Delta}(x) &= \frac{1}{(2\pi)^4}\PV\int \frac{e^{ik_\mu x_\mu}}{k_\lambda^2+\kappa_0^2}(dk)\quad .\label{eq:AppDeltaBarIntegralEqn}
\intertext{At this point we need the integral representation,}
\PV\left[\frac{1}{\tau}\right] &= -\frac{i}{2}\int_{-\infty}^{\infty} e^{ia\tau}\frac{a}{|a|}da\quad .\label{eq:AppPVIntegralEqn}
\end{align}
Here is a proof:
\begin{align*}
\hat{f}(k) &= \int e^{ikx}f(x)dx\\
f(x) &= \frac{1}{2\pi}\int e^{-ixk}\hat{f}(k)dk = \frac{1}{2\pi}\int e^{ixk}dk \int e^{ikx'}f(x')dx'\\
\PV\left[\frac{1}{\tau}\right] &= \frac{1}{2\pi}\int e^{-i\tau\alpha} d\alpha \left[ \PV\int e^{i\alpha\tau'}\frac{1}{\tau'}d\tau'\right]\quad .
\end{align*}
By contour integration it is easy to prove that
\begin{align*}
\PV\int_{-\infty}^\infty e^{i\tau'\alpha}\frac{1}{\tau'}d\tau' &= \left\lbrace \begin{array}{lr}
i\pi, &\quad \alpha>0\\
-i\pi, &\quad \alpha <0
\end{array}\right\rbrace=i\pi\frac{\alpha}{|\alpha|}\quad .
\intertext{Then we obtain}
\PV \left[\frac{1}{\tau}\right] &= \frac{i}{2}\int_{-\infty}^\infty e^{-i\alpha\tau}\frac{\alpha}{|\alpha|}d\alpha\quad .
\intertext{A change $\beta := -\alpha$ gives us}
\PV\left[\frac{1}{\tau}\right] &= \frac{i}{2}\int_{\infty}^{-\infty} e^{i\beta\tau}\frac{\beta}{|\beta|}d\beta = -\frac{i}{2}\int_{-\infty}^\infty e^{i\beta\tau}\frac{\beta}{|\beta|}d\beta\\
&= \frac{1}{2i}\int_{-\infty}^\infty e^{i\alpha\tau} \frac{\alpha}{|\alpha|}d\alpha.\qquad \square
\intertext{With this information we obtain:}
\bar{\Delta}(x) &= -\frac{i}{2(2\pi)^4}\int_{-\infty}^\infty \frac{a}{|a|} \int e^{ia\tau +ik_\mu x_\mu}(dk)da\  ,
\end{align*}
and with $\tau = k_\lambda^2+\kappa_0^2$, we find
\begin{align}
\bar{\Delta}(x) &= -\frac{i}{2(2\pi)^4}\int_{-\infty}^\infty \frac{a}{|a|}\int e^{iak_\mu^2+ik_\mu x_\mu}e^{ia\kappa_0^2}(dk)da\notag\\
&= \frac{1}{32\pi^2}\int_{-\infty}^\infty \frac{1}{a^2} e^{-i\frac{x_\mu^2}{4a}+ i a \kappa_0^2} da\quad ,\label{eq:AppDeltaBarIntegralEqn2}
\end{align}
where we made use of the formula
\begin{equation}
\int e^{iak_\mu^2+ik_\mu x_\mu}(dk) = \int e^{iak_\mu^2}e^{-i\frac{x_\mu^2}{4a}}(dk) = i\frac{\pi^2}{a|a|}e^{-i\frac{x_\mu^2}{4a}}\quad .\label{eq:AppIntegralTransform}
\end{equation}
The new variable $\alpha:=\tfrac{1}{4a}$ brings \eqref{eq:AppDeltaBarIntegralEqn2} into the form
\begin{equation}
\bar{\Delta}(x) = \frac{1}{8\pi^2}\int e^{i\lambda \alpha +i\frac{\kappa_0^2}{4\alpha}} d\alpha\  ,\quad \lambda=-x_\mu^2\quad ,
\end{equation}
which can be rewritten as:
\begin{equation}
\bar{\Delta}(x) = \bar{\Delta}(\lambda) = \frac{1}{4\pi^2}\int_0^\infty \cos\left(\lambda\alpha + \frac{\kappa_0^2}{4\alpha}\right)d\alpha
= \frac{1}{4\pi^2}\frac{\partial}{\partial\lambda}\int_0^\infty \sin\left(\lambda\alpha + \frac{\kappa_0^2}{4\alpha}\right)\frac{d\alpha}{\alpha}\  .
\label{eq:AppDeltaBarLambdaIntegroDiffEqn}
\end{equation}
In order to evaluate the last integral we define a new variable of integration:
\begin{equation}
\alpha:=\tfrac{\kappa_0}{2\sqrt{|\lambda|}}e^\vartheta\quad .\label{eq:AppAlphaIntegralVariableTransform}
\end{equation}
Therefore the last integral in \eqref{eq:AppDeltaBarLambdaIntegroDiffEqn} turns into
\begin{align}
\int_0^\infty \sin\left(\lambda\alpha + \frac{\kappa_0^2}{4\alpha}\right)\frac{d\alpha}{\alpha} &= \int_{-\infty}^\infty \sin\left[ \frac{\kappa_0\sqrt{|\lambda|}}{2}
\left( \frac{\lambda}{|\lambda|} e^\vartheta + e^{-\vartheta}\right) \right]d\vartheta\notag\\
&= \begin{cases}
\int_{-\infty}^\infty \sin\left(\kappa_0\lambda^{\frac{1}{2}}  \cosh\vartheta\right)d\vartheta & \lambda>0\\
\int_{-\infty}^\infty \sin\left(\kappa_0(-\lambda)^{\frac{1}{2}}  \sinh\vartheta\right)d\vartheta & \lambda<0
\end{cases}\notag\\
&=\begin{cases}
\pi J_0(\kappa_0\lambda^{\frac{1}{2}}) & \lambda>0\\
0 & \lambda<0\quad .
\end{cases}\label{eq:AppSineTransformToBessel}
\end{align}
This discontinuous value can be compactly represented by
\begin{equation}
\int_0^\infty \sin\left(\lambda\alpha + \frac{\kappa_0^2}{4\alpha}\right)\frac{d\alpha}{\alpha} = \pi \mathfrak{Re}\left[H_0^{(1)}(\kappa_0\lambda^{\frac{1}{2}})\right]\quad ,\label{eq:AppCompactReprOfDiscontinuity}
\end{equation}
provided $\sqrt{\lambda}$ with $\lambda<0$ is interpreted as $i\sqrt{|\lambda|}$. Finally,
\begin{equation}
\bar{\Delta}(x) = \frac{1}{4\pi}\delta(\lambda) -\frac{\kappa_0^2}{8\pi}\mathfrak{Re}\left[\frac{H_1^{(1)}(\kappa_0\lambda^{\frac{1}{2}}) }{\kappa_0\lambda^{\frac{1}{2}}}\right]\quad ,\label{eq:AppDeltaBarCompact}
\end{equation}
where
\begin{equation}
\mathfrak{Re}\left[\frac{H_1^{(1)}(\kappa_0\lambda^{\frac{1}{2}}) }{\kappa_0\lambda^{\frac{1}{2}}}\right] = \begin{cases}
\frac{J_1(\kappa_0\lambda^{\frac{1}{2}} )}{\kappa_0\lambda^{\frac{1}{2}}} & \lambda >0\\
0 & \lambda <0\quad \text{space-like,}
\end{cases}\label{eq:AppRealPartHankelFunctionSpaceLikeParameter}
\end{equation}
and the delta function of $\lambda$ arises from the discontinuity of \eqref{eq:AppCompactReprOfDiscontinuity} at $\lambda=0$. Clearly $\bar{\Delta}(x)$, and therefore the Pauli-Jordan function
$\Delta(x)$ vanishes if $-x^2<0$, which is one of the defining properties of the latter function.
Setting $\kappa_0=0$, we obtain:
\begin{equation}
\bar{D}(x) = -\frac{1}{2}D(x)\epsilon(x)=\frac{1}{4\pi}\delta(\lambda) = \frac{1}{4\pi}\delta(x_\mu^2)\quad .\label{eq:AppDeltaBarAsDiracDelta}
\end{equation}
An integral representation for $\Delta(x)$ itself can be constructed directly with the aid of the inverse Fourier transform of \eqref{eq:AppPVIntegralEqn}. 
We found that
\begin{equation*}
2i\PV\left[\frac{1}{\tau}\right] = \int e^{ia\tau}\frac{a}{|a|}da\quad .
\end{equation*}
So the Fourier transform of $\tfrac{a}{|a|}$ is given by
\begin{align*}
\frac{a}{|a|} &= \frac{2i}{2\pi}\int e^{-ia\tau}\frac{1}{\tau}d\tau\quad .
\intertext{Let $\tau=-\beta$:}
\frac{a}{|a|} = \frac{i}{\pi}\int e^{ia\beta}\frac{1}{\beta}d\beta &= -\frac{i}{\pi} \int e^{ia\beta} \frac{1}{\beta}d\beta = -\frac{i}{\pi}\PV\int e^{ia\tau}\frac{1}{\tau}d\tau\ .
\end{align*}
So we can write
\begin{equation}
\epsilon(x) = -\frac{\epsilon_\mu x_\mu}{|\epsilon_\mu x_\mu|}=\frac{i}{\pi}\PV\int e^{i\epsilon_\mu x_\mu\tau} \frac{1}{\tau}d\tau\quad .
\end{equation}
On employing the first expression of \eqref{eq:AppDeltaBarIntegralEqn2} for $\bar{\Delta}(x)$, we obtain ($\epsilon^2(x)=1$):
\begin{equation}
\Delta(x) = 2\bar{\Delta}(x)\frac{\epsilon_\mu x_\mu}{|\epsilon_\mu x_\mu|} = -\frac{2}{(2\pi)^5}\int(dk)\int\frac{a}{|a|}da\  \PV\int e^{i(k_\mu+\epsilon_\mu\tau)x_\mu}e^{ia(k^2_\mu+\kappa_0^2)}\frac{1}{\tau}d\tau\  .
\end{equation}
With the transformation $k_\mu\rightarrow k_\mu-\epsilon_\mu\tau$ we obtain:
\begin{equation}
\Delta(x) = -\frac{2}{(2\pi)^5}\int(dk)\int \frac{a}{|a|}da\ \PV\int \frac{1}{\tau}e^{-2ia\epsilon_\mu k_\mu\tau}e^{ia\epsilon^2_\mu\tau^2}e^{ik_\mu x_\mu}e^{ia(k_\mu^2+\kappa_0^2)}d\tau\  .
\label{eq:AppDeltaIntegralEqnWithPV}
\end{equation}
Since $\epsilon_\mu$ is a time-like vector with $\epsilon_0>0$ we can argue that this equation is independent of $\epsilon_\mu$. This characteristic can be maintained with $-\epsilon_\mu^2$ an
arbitrarily small positive number. We can therefore evaluate \eqref{eq:AppDeltaIntegralEqnWithPV} in the limit $\epsilon_\mu^2\rightarrow 0$. Then, with
\begin{align}
\PV\int e^{-2ia\epsilon_\mu k_\mu\tau}\frac{1}{\tau}d\tau &= -i\pi\frac{a}{|a|}\frac{\epsilon_\mu k_\mu}{|\epsilon_\mu k_\mu|} = i\pi\frac{a}{|a|}\epsilon(k)\quad ,\label{eq:AppEpsilonOfKPVEqn}
\intertext{we find:}
\Delta(x) &= -\frac{i}{(2\pi)^4}\int(dk)\int e^{ia(k_\mu^2+\kappa_0^2)}e^{ik_\mu x_\mu}\epsilon(k)da\notag\\
 &= -\frac{i}{(2\pi)^3}\int e^{ik_\mu x_\mu}\delta(k_\mu^2+\kappa_0^2)\epsilon(k)(dk)\quad .\label{eq:AppDeltaIntegralEqnR3,1}
\end{align}
Formula \eqref{eq:AppEpsilonOfKPVEqn} uses the step function $\epsilon(\alpha)$ in the form
\begin{equation*}
\epsilon(\alpha)=\frac{1}{i\pi}\PV\int e^{i\alpha\tau}\frac{1}{\tau}d\tau = \Theta(\alpha)-\Theta(-\alpha) = \begin{cases}
1 & \alpha>0\\
-1 & \alpha<0\quad .
\end{cases}
\end{equation*}
The result \eqref{eq:AppDeltaIntegralEqnR3,1} makes it evident that $\Delta(x)$ satisfies the proper homogeneous differential equation ($x\delta(x)=0$):
\begin{equation*}
\left(\partial^2-\kappa_0^2\right) \Delta(x) = 0\quad .
\end{equation*}
An integral representation for $\Delta^{(1)}(x)$ can be obtained immediately from that of $\Delta(x)$.

According to Schwinger's definition and \eqref{eq:AppDeltaIntegralEqnR3,1}:
\begin{align}
\Delta^{(1)}(x) &= \frac{1}{\pi}\PV\int\Delta(x-\epsilon\tau)\frac{1}{\tau}d\tau \notag\\
&= -\frac{i}{\pi}\frac{1}{(2\pi)^3}\int(dk)\epsilon(k)
\underset{{i\pi\epsilon(k)}}{\underbrace{\left[\PV\int\frac{1}{\tau}e^{-ik_\mu\epsilon_\mu\tau}d\tau\right] }} e^{ik_\mu x_\mu}\delta(k^2+\kappa_0^2)\notag\\
&\overset{\epsilon^2(k)=1}{=} \frac{1}{(2\pi)^3}\int e^{ik_\mu x_\mu}\delta(k^2+\kappa_0^2)(dk)\quad .\label{eq:AppDelta1IntEqR3,1}
\end{align}
Again, it is evident that $\Delta^{(1)}(x)$, like $\Delta(x)$, satisfies $(\partial^2-\kappa_0^2)\Delta^{(1)}=0$.
Of course, we can also evaluate $\Delta^{(1)}(x)$ in a manner similar to $\bar{\Delta}(x)$. We employ the integral representation for $\delta(k^2+\kappa_0^2)$ in \eqref{eq:AppDelta1IntEqR3,1}.
Setting $\tau=k_\mu^2+\kappa_0^2$ in
\begin{equation*}
\delta(\tau) = \frac{1}{2\pi}\int e^{ia\tau}da
\end{equation*}
and performing the integration over $k$ space, we obtain:
\begin{align}
\Delta^{(1)} (x) &= \frac{1}{(2\pi)^4}\int da\int e^{iak_\mu^2+ik_\mu x_\mu}e^{ia\kappa_0^2} (dk)\notag\\
&= \frac{i}{16\pi^2}\int e^{i\frac{x_\mu^2}{4a}+ia\kappa_0^2}\frac{a}{|a|}\frac{da}{a^2}\notag\\
&= \frac{1}{4\pi^2}\int e^{i\lambda\alpha + i \frac{\kappa_0^2}{4\alpha} }\frac{\alpha}{|\alpha|}d\alpha\quad .\label{eq:AppDelta1IntEqRAlpha}
\end{align}
Rewriting \eqref{eq:AppDelta1IntEqRAlpha} yields
\begin{align}
\Delta^{(1)}(x) = \Delta^{(1)}(\lambda) &= -\frac{1}{2\pi^2}\int_0^\infty \sin(\lambda\alpha + \frac{\kappa_0^2}{4\alpha})d\alpha\\
&= \frac{1}{2\pi^2}\frac{\partial}{\partial \lambda}\int_0^\infty \cos(\lambda\alpha +\frac{\kappa_0^2}{4\alpha})\frac{1}{\alpha}d\alpha\  .
\end{align}
Again, using transformation \eqref{eq:AppAlphaIntegralVariableTransform}, we find:
\begin{align}
\int_0^\infty \cos\left(\lambda\alpha+\frac{\kappa_0^2}{4\alpha}\right)\frac{1}{\alpha}d\alpha &= 
\begin{cases}
\int_{-\infty}^\infty \cos\left( \kappa_0\lambda^{\scriptsize \frac{1}{2}} \cosh\vartheta \right)d\vartheta & \lambda>0\\
\int_{-\infty}^\infty \cos\left( \kappa_0(-\lambda)^{\scriptsize \frac{1}{2}} \sinh\vartheta \right)d\vartheta & \lambda<0
\end{cases}\notag\\
&= \begin{cases}
-\pi N_0 \left(\kappa_0 \lambda^{\scriptsize \frac{1}{2}}\right) & \lambda>0\\
2K_0\left( \kappa_0(-\lambda)^{\scriptsize \frac{1}{2}}\right) & \lambda <0\quad ,
\end{cases}
\end{align}
which is summarized in
\begin{equation}
\int_0^\infty \cos\left( \lambda\alpha +\frac{\kappa_0^2}{4\alpha} \right)\frac{d\alpha}{\alpha} = -\pi\  \mathfrak{Im}\left[ H_0^{(1)} \left( \kappa_0 \lambda^{\scriptsize \frac{1}{2}}\right)\right]\quad .
\end{equation}
Unlike \eqref{eq:AppSineTransformToBessel} there is no discontinuity at $\lambda = 0$. Therefore,
\begin{equation}
\Delta^{(1)}(x) = \frac{\kappa_0^2}{4\pi}\  \mathfrak{Im}\left[ \frac{H_1^{(1)}\left( \kappa_0 \lambda^{\scriptsize \frac{1}{2}}\right) }{ \kappa_0 \lambda^{\scriptsize \frac{1}{2}}}
\right] = \begin{cases}
\frac{\kappa_0^2}{4\pi} \frac{N_1\left( \kappa_0 \lambda^{\scriptsize \frac{1}{2}}\right)}{\kappa_0 \lambda^{\scriptsize \frac{1}{2}}} & \lambda >0\\
\frac{\kappa_0^2}{2\pi^2} \frac{K_1 \left( \kappa_0 (-\lambda)^{\scriptsize \frac{1}{2}}\right)}{ \kappa_0 (-\lambda)^{\scriptsize \frac{1}{2}}} & \lambda <0\  .
\end{cases}\label{eq:AppDelta1FinalRepresentaion}
\end{equation}
The singularity of $\Delta^{(1)}(x)$ at $\lambda=0$ shows up by writing
\begin{align}
\Delta^{(1)}(x) &= -\frac{1}{2\pi^2\lambda}+\frac{\kappa_0^2}{4\pi}\mathfrak{Im}\left[ \frac{H_1^{(1)} \left( \kappa_0 \lambda^{\scriptsize \frac{1}{2}}\right) }{ \kappa_0 \lambda^{\scriptsize \frac{1}{2}} } +\frac{2i}{\pi}\frac{1}{\kappa_0^2\lambda} \right]
\intertext{and, on letting $\kappa_0\rightarrow 0$, we obtain:}
D^{(1)}(x) &= -\frac{1}{2\pi^2\lambda} = \frac{1}{2\pi^2} \frac{1}{x_\mu^2}\quad .
\end{align}

\bibliography{ArticleReferences.bbl}

\begin{thebibliography}{Tom97}

\bibitem[BH34]{Bethe1934}
H.~Bethe and W.~Heitler.
\newblock On the stopping of fast particles and on the creation of positive
  electrons.
\newblock {\em Proceedings of the Royal Society of London}, 146, August 1934.

\bibitem[BHJ25]{Born1925}
Max Born, Werner Heisenberg, and Paul Jordan.
\newblock \"{U}ber quantentheoretische umdeutung kinematischer und mechanischer
  beziehungen (en. quantum theoretical re-interpretation of kinematic and
  mechanical relations).
\newblock {\em Zeitschrift f\"{u}r Physik}, 35:557--615, 1925.

\bibitem[DG00]{Dittrich2000}
Walter Dittrich and Holger Gies.
\newblock {\em Probing the Quantum Vacuum\\ Perturbative Effective Action
  Approach in Quantum Electrodynamics and its Application}.
\newblock Springer, 2000.

\bibitem[Dit14]{Dittrich2014}
Walter Dittrich.
\newblock On schwinger's formula for pair production.
\newblock {\em International Journal of Modern Physics A},
  29(13):1430033--(1--8), May 2014.

\bibitem[DR85]{Dittrich1985}
Walter Dittrich and Martin Reuter.
\newblock {\em Effective Lagrangians in Quantum Electrodynamics}.
\newblock Lecture Notes in Physics. Springer, 1985.

\bibitem[Esp07]{Esposito2007}
S.~Esposito.
\newblock {\em Annalen der Physik}, 16, 2007.

\bibitem[PW34]{Pauli1934}
W.~Pauli and V.~Weisskopf.
\newblock {\em Helvetica Physica Acta}, 7, 1934.

\bibitem[Sch49]{Schwinger1949}
J.~Schwinger.
\newblock Quantum electrodynamics. ii. vacuum polarization and self-energy.
\newblock {\em Physical Review}, 75:651--679, February 1949.

\bibitem[Tom97]{Tomonaga1997}
Shin'ichir\={o} Tomonaga.
\newblock {\em The Story of Spin}.
\newblock The University of Chicago Press, 1997.

\end{thebibliography}

\end{document}